\documentstyle[aps,pra,epsfig,twocolumn]{revtex}

\def\be{\begin{equation}}
\def\ee{\end{equation}}
\def\bea{\begin{eqnarray}}
\def\eea{\end{eqnarray}}
\def\bma{\begin{mathletters}}
\def\ema{\end{mathletters}}
\def\C{\hbox{$\mit I$\kern-.7em$\mit C$}}
\newcommand{\eins}{\mbox{$1 \hspace{-1.0mm}  {\bf l}$}}

\begin{document}
\draft

\title{Classification of multi--qubit mixed states: separability and distillability properties}

\author{W. D\"ur$^1$ and J. I. Cirac$^1$}

\address{$^1$Institut f\"ur Theoretische Physik, Universit\"at Innsbruck,
A-6020 Innsbruck, Austria}

\date{\today}

\maketitle

\begin{abstract}
We give a complete, hierarchic classification for arbitrary
multi--qubit mixed states based on the separability properties of
certain partitions. We introduce a family of $N$--qubit states to
which any arbitrary state can be depolarized. This family can be
viewed as the generalization of Werner states to multi--qubit
systems.  We fully classify those states with respect to their
separability and distillability properties. This provides
sufficient conditions for nonseparability and distillability for
arbitrary states.
\end{abstract}


\narrowtext

% --------------------------------------------------------------
% Introduction
%---------------------------------------------------------------

\section{Introduction}

Entanglement is one of the basic concepts of Quantum Mechanics and an important 
feature of most applications of Quantum Information. It arises when a state of a 
multiparticle quantum system cannot be prepared by acting on the particles 
individually, i.e. is non--separable. Despite of the fact that we do not know 
yet how to classify and quantify entanglement in general, much progress was made 
in recent years. In particular, the concept of entanglement distillation or 
purification\cite{Be96} was introduced. This process, which is the creation of 
(few) maximally entangled states out of many not--maximally entangled ones, 
turned out to be one of the most important concepts in quantum information 
theory. When combined with teleportation \cite{Be93}, it allows to send quantum 
information over noisy channels \cite{Be93,Noisy} and to convey secret 
information via quantum privacy amplification \cite{De96}.

Particular important states of two qubits are the so called Werner states (WS) 
\cite{We89}, which are mixtures of a maximally entangled state, e.g. 
$|\Phi^+\rangle=\frac{1}{\sqrt{2}}(|00\rangle+|11\rangle)$, with the totally 
depolarized state. These states are fully characterized by the fidelity $F$, 
which measures the overlap of the maximally entangled state $|\Phi^+\rangle$ 
with the WS. They play an essential role in the understanding of the 
entanglement and distillability properties of two qubit systems\cite{Wo98}. On 
the one hand, it has been shown that WS are separable for $F \leq 1/2$ and 
non--separable (entangled) for $F>1/2$. On the other hand, Bennett {\it et al.} 
\cite{Be96} showed that one can purify WS with arbitrary high fidelity out of 
many pairs with $F>1/2$ by using local operations and classical communication. 
Furthermore, any arbitrary state can be depolarized to a WS without changing the 
fidelity $F$, which automatically provides a sufficient criterion for 
non--separability \cite{Pe96,Ho96,Ho97} and distillability \cite{Ho97b} for 
arbitrary states.

The description of the entanglement and distillability properties of systems 
with more than two particles is still almost unexplored (see Refs. 
\cite{Mu98,multi}, however). In \cite{Du99}, some steps towards the 
understanding of 3--particle entanglement of mixed states were taken. In 
particular, a complete classification of arbitrary 3--qubit states was proposed 
and the distillability and separability properties of a family of states was 
obtained. In this paper we generalize the ideas introduced in \cite{Du99} to 
multiparticle quantum systems. We provide a complete classification of a family 
of states of $N$--qubit systems. These states are characterized in terms of 
$2^{N-1}$ parameters and play the role of WS in such systems, since any 
arbitrary state can be depolarized to this form. We fully analyze the 
separabilty and distillability properties of this family, thereby generalizing 
the purification procedure introduced in \cite{Du99} to multiqubit systems. This 
automatically provides us - as in the bipartite case - with sufficient 
conditions for arbitrary multi--qubit states. Among other things, this allows us 
to give the necessary and sufficient separability and distillability conditions 
of mixtures of a maximally entangled state and the completely depolarized state 
\cite{Br99,Vi99}. Furthermore we introduce a hierarchic classification of 
general $N$--qubit states with respect to their entanglement properties.

The paper is organized as follows. We start by briefly reviewing some of the 
present knowledge about distillability and entanglement of bipartite quantum 
systems in Sec. \ref{2x2}. In the following we generalize the results of 
\cite{Du99} to multi--qubit systems. We start out by giving a classification of 
arbitrary $N$--qubit systems in Sec. \ref{Secmulti}. Then we introduce a family 
of states that can be obtained via depolarization from an arbitrary one in Sec. 
\ref{family}. Here we also investigate the separability and distillability 
properties of this family. Sec. \ref{Examples} gives examples to illustrate the results 
obtained in the previous Sections. In particular, we analyze in detail the 
simplest cases of 3 and 4 qubit systems. In Sec. \ref{mixture}, we apply our 
results to the case where we have a maximally entangled state of $N$ qubits 
mixed with the totally depolarized state. Finally we conclude and summarize in 
Sec. \ref{Summary}.

%--------------------------------------------------------------------------------
%               Bipartite systems and PT
%--------------------------------------------------------------------------------

\section{Bipartite systems and partial transposition}\label{2x2}

Let us start out by briefly reviewing the separability and distillability
properties of bipartite systems. A bipartite mixed state $\rho$ is called separable iff
it can be prepared locally, i.e. it can be written as a convex combination of
(unnormalized) product states
\be
\rho = \sum_i |a_i\rangle_{\rm party1}\langle a_i|
  \otimes |b_i\rangle_{\rm party2}\langle b_i|.
\ee A state is called distillable iff one can create out of
(infinitely) many copies of $\rho$ one maximally entangled state,
e.g. $|\Phi^+\rangle$. In practice, it is difficult to
decide whether a given state is separable or distillable
respectively. As shown by Peres\cite{Pe96} and the
Horedecki\cite{Ho96,Ho97,Ho97b}, the partial transposition of a
density operator turns out to provide a simple, sufficient
criterion for the classification of bipartite systems. Given an
operator $X$ acting on $\C^{d_1}\otimes\C^{d_2}$, the partial
transposition with respect to the first subsystem in the standard
basis $\{|1\rangle,|2\rangle,\ldots,|d_1\rangle\}$, $X^{T_A}$, is
defined as follows: \be X^{T_A} \equiv \sum_{i,j=1}^{d_1} \langle
i|X|j\rangle \; |j\rangle\langle i|. \ee Clearly, the partial
transposition of the operator $X$ is basis dependent, but the
eigenvalues are not. We say that a self adjoint operator has
positive partial transposition ($X^{T_A} \geq 0$) - PPT - iff all
eigenvalues of $X^{T_A}$ are non--negative. On the opposite, we
say an operator has non--positive partial transposition (NPPT) iff
at least one eigenvalue is negative. Sometimes NPPT is also called
''negative partial transposition'' (NPT).

For bipartite two--level systems ($d_1=d_2=2$) it was shown that positive 
partial transposition (PPT) is a necessary and sufficient condition for 
separability\cite{Pe96,Ho97} while negative partial transposition (NPT) is a 
necessary and sufficient condition for distillability\cite{Ho97b}. For higher 
dimensional systems, however, the partial transposition only provides necessary 
conditions for separability\cite{Ho97} and it seems that it provides only a 
necessary condition for distillability\cite{Du99b,Di99a}. In $\C^2\otimes\C^d (d 
\geq 2)$ systems we have that a sufficient condition for separability is that 
$\rho=\rho^{T_A}$ \cite{Le99}, while the negativity of the partial transposition 
already ensures distillability of those systems\cite{Du99b}.

In the following we generalize the notion of separability and distillability to
multi--qubit systems. It turns out that in order to characterize an important
family of multi--qubit mixed states, it is useful to consider bipartite splits of
multiparticle systems and their corresponding partial transpositions. Since a
bipartite split of a multi--qubit system can be viewed as a state in
$\C^{d_1}\otimes\C^{d_2}$, the partial transposition of the density operator
$\rho$ is well defined in this case.

%--------------------------------------------------------------------------------
%               N-qubit systems
%--------------------------------------------------------------------------------
\section{Multi--qubit systems:}\label{Secmulti}

We will give a classification of general $N$--qubit systems in
terms of the separability properties of their partitions. In
particular, we consider $k$--partite splits (that are partitions
dividing a $N$--partite systems into $k\leq N$ parties), which
gives rise to a whole hierarchy of classes. 

%----------------------------------------------------

\subsection{Separability with respect to certain splits}

\begin{figure}[ht]
\begin{picture}(230,110)
\put(0,5){\epsfxsize=200pt\epsffile[62 648 246 758]{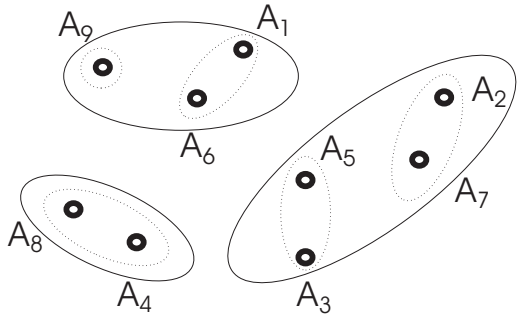}}
\end{picture}
\caption[]{Example of two partitions of a 9--qubit system into three sets
(full line - $S_3=(A_1A_6A_9)-(A_2A_3A_5A_7)-(A_4A_8)$) and five sets (dotted line - $S_5=(A_1A_6)--(A_9)-(A_2A_7)-(A_3A_5)-(A_4A_8)$). We have that $S_5$ is contained in $S_3$.} \label{Fig2}
\end{figure}

Let us start by generalizing the notion of separability to the case of
multiparticle systems. We consider $N$ parties, each holding a system with
dimension $d_i$, i.e. ${\cal H}=\C^{d_1}\otimes \ldots \otimes\C^{d_N}$. We call
$\rho$ fully separable if it can be written as a convex combination
of (unnormalized) product states, i.e. 
\be \rho = \sum_i |a_i\rangle_{\rm
party1}\langle a_i| \otimes |b_i\rangle_{\rm party2}\langle b_i|\otimes \ldots
\otimes |n_i\rangle_{{\rm party} N}\langle n_i|.\label{sepa1} 
\ee 
In the following, we will consider a system of $N$ qubits, each hold by one of the 
parties $A_1,A_2,\ldots,A_N$. In this case, $d_1=d_2=\ldots d_N=2$. Let us now 
consider a partition of the $N$--qubit system into $k \leq N$ sets, which we 
call a $k$--partite split of the system (see Fig. \ref{Fig2}). That is we allow 
some of the parties to act together such that finally $k$ parties remain.  As a 
special case, we have 2--partite splits which we will also call bipartite 
splits. A state $\rho$ is called $k$--separable with respect to this specific 
partition (or equivalently split) iff it is fully separable in the sense that we 
consider $\rho$ as a $k$--party system, i.e. as a state in ${\cal 
H}=\C^{d_1}\otimes \ldots \otimes\C^{d_k}$.

Considering all possible partitions (including all permutations)
of the $N$--qubit system and determining the corresponding
separability properties is sufficient to fully characterize the
system in terms of its entanglement properties. However, the
number of possible partitions grows rapidly with the number
of parties involved (see Sec. \ref{Partitions}), and it turns out
that the information given by all this properties is redundant in
some cases. We thus propose a hierarchic classification in terms
of the separability properties with respect to the partitions,
i.e. we consider all $k$--partite splits at level $k$ of our
classification. This turns out to be useful, since in some cases
(see Sec. \ref{family}) the information of one level (in
particular level 2) already implies all properties at the other
levels. Furthermore, there are connections between the different
levels, which can be used to reduce the effort to determine the
full entanglement properties of the system. However, we learned in
the case of 3--qubits that these connections are sometimes not
obvious or even counter--intuitive. For example we have that for a
3--qubit system separability with respect to all bipartite
splits (i.e. partitions into two sets) is not sufficient to
guarantee 3-separability (i.e. full separability when considering
each system $A_1$,$A_2$,$A_3$ as a separate party) of the system
\cite{Du99,Be98}.

\subsection{Classification of arbitrary states $\rho$}\label{Classification}

In \cite{Du99}, the biseparability properties of the state $\rho$ where used to
classify those states completely, and all together (apart from permutations
among the parties) 5 distinct classes were found (see Sec.\ref{Example3} for
details). While splits of the system into two or three parties were sufficient
to fully classify three qubit systems, it turns out that for $N$--qubit systems
we need all possible partitions of the system for a complete classification.

\subsubsection{Hierarchic classification}

The basic idea of the hierarchic classification we propose here is
to consider all possible $k$--partite splits of a $N$--partite
system for all ($k \in \{N, N-1, \ldots ,2\}$) and determine for
each split whether it is $k$--separable or not. For simplicity, we
divide this procedure into levels, starting with $k=N$, continue
with $k=N-1$ etc. until we reach $k=2$. This minimizes the
necessary effort for a full classification, since, as mentioned
in the previous section, there are connections between the
different levels which will be explained in more detail in Sec.
\ref{Contained}.

Level $k$ of the characterization consists of the complete determination of the
$k$--separability properties of the state $\rho$, i.e. considering all possible
partitions into exactly $k$ sets and determine whether the state is separable.
At each level $k$, we have various classes, namely all possible combinations of
$k$--separability and $k$--inseparability. If the number of $k$-partite splits
is $k_0$, we have $2^{k_0}$ possible configurations at this level in principle.

However, the different levels of this structure are not independent of each
other and thus some of the possible configurations are forbidden by the
structure at higher/lower levels. We call each allowed configuration of the
whole hierarchic classification a ''class'', since it corresponds to different
physical properties. We have that permutations of the parties lead to different
classes.

Note that all levels of this characterization are required to fully classify a
state. It is not sufficient to give only the number of $k$--separable
splits at each level and define classes in terms of this numbers, as done for
$N=3$ in \cite{Du99}. In this case, one obtains the  remaining configurations by
permuting the parties, while for $N > 3$ this last property is no longer valid,
i.e. one can have two physically different situations (not only up to
permutations) corresponding to the same number of $k$--separable states at a
certain level. This will be explained in more detail in Sec.\ref{Example4}.

In principle, it may turn out that some of the classes we give here are empty. In
fact, for the family of states we are going to consider in the following, we
have that $k$--separability is implied by the corresponding 2--separabilities,
i.e. by the biseparibility properties of all bipartite splits containing the
$k$--partite split $S_k$ in question. This means that these states are already
fully classified by the structure at level 2 (biseparability properties).
However, for $N=3$ examples for 3--inseparable states which are biseparable with
respect to all bipartite splits are known \cite{Be98}, which makes it likely
that similar examples (apart from the trivial generalization of those states
$\rho_B$ to $N$ qubits by taking e.g. $|0\rangle\langle 0|^{\otimes N-3}\otimes
\rho_B$) also exist for $N > 3$.

\subsubsection{Partitions of $N$--qubits}\label{Partitions}

A partition of $N$, $\cal{L}$, is given by ${\cal L}_{\vec r}=\{1^{r_1}2^{r_2}
\ldots N^{r_N}\}$ with $\sum_{j=0}^N j r_j = N$, and the number of
sets\footnote{The number of sets is equivalent to the number of parties} $k=\sum
r_i$. For example, $N=4$ and ${\cal L}_{21}=\{1^2,2^1\}$ denotes all possible
partitions of $N$ into 3 sets such that one set consists of 2 parties, the other
two sets consist of one party each. Note that we may have many partitions which
correspond to the same number of sets, say $k$, which we called ''$k$--partite
splits''.

Using the well developed theory of partitions (see e.g. \cite{An}), one finds
that the number of possible configurations (including all possible permutations
among the parties) for a certain partition ${\cal L}_{\vec r}$ is given by
\be
|{\cal L}_{\vec r}|=\frac{N!}{\prod r_j! \prod (j!)^{r_j}}
\ee
The total number of partitions ${\cal L}_{\vec r}$ is given by the partition
function $p(N)$, which grows rapidly with $N$, e.g. $p(10)$=42,
$p(50)$=204226, $p(100)$=190569292. A closed expression for $p(N)$ is known and
can be found e.g. in \cite{An}. Using these relations, one can in principle obtain the number
of possible $k$--partite splits of a $N$--qubit system.

\subsubsection{Contained splits and implications for classification}\label{Contained}

\begin{figure}[ht]
\begin{picture}(230,110)
\put(0,5){\epsfxsize=230pt\epsffile[1 594 594 821]{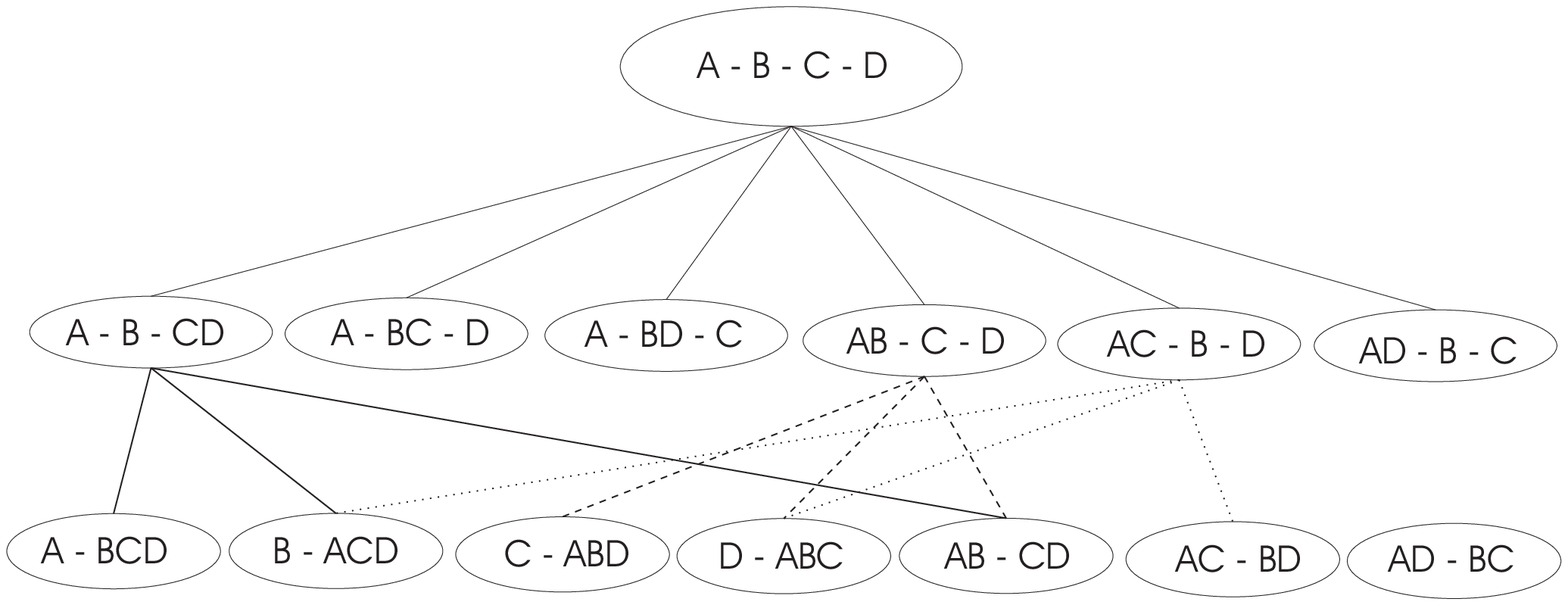}}
\end{picture}
\caption[]{Hierarchy structure for classification of $4$--qubit
systems. Note that not all connections at the lowest level are
drawn.} \label{Fig1}
\end{figure}

Let $l < k$. We say a $k$--partite split $S_k$ {\it belongs to} (equivalently
{\it is contained in}) a $l$--partite split $S_l$ iff $S_l$ can be obtained from
$S_k$ by joining some of the parties of $S_k$. For 3 parties $(N=3)$ we have, for
example, that the bipartite split $A_1-(A_2A_3)$ contains the 3--partite split
$A_1-A_2-A_3$, since the split $A_1-(A_2A_3)$ can be obtained by joining the
parties $(A_2A_3)$. Note that in general we do not have a one--to--one
correspondence in neither direction. On one hand, each $k$--partite split is
contained in various $l$--partite splits, while on the other hand a number of
different $k$--partite splits may be contained in the same $l$--partite
split (see also Fig.\ref{Fig1}).

Furthermore, $k$--separability with respect to a certain $k$--partite split
$S_k$ implies $l$--separability with respect to all those $l$--partite splits
which contain $S_k$ ($l < k$). However we learned in the case of three qubits
\cite{Du99} that 2--separability with respect to all possible 2--partite splits
is not sufficient to guarantee the corresponding 3--separability in general.
Thus we have that $l$--separability with respect to all those $l$--partite
splits which contain a certain $k$--partite split $S_k$ is a necessary, but not
sufficient condition for $k$--separability with respect to $S_k$.

Let us apply this observation to our classification. We have that
$k$--separability partly fixes the structure at lower levels $l<k$
($k$--inseparability has no influence at lower levels), while
$l$--inseparability fixes some properties at higher levels $k>l$
($l$--separability only provides necessary conditions for
separability at higher levels). In particular, $k$--separability
with respect to a certain $k$--partite split $S_k$ already implies
$l$--separability of all $l$--partite splits containing $S_k$ ($l
< k$). On the other hand, $l$--inseparabilty with respect to a
certain $l$--partite split $S_l$ implies that all $k$--partite
splits which belong to $S_l$ ($k>l$) are also $k$--inseparable.
This means that once one finds a $k$--partite split $S_k$ to be
$k$--separable, one does not have to consider all $l$--partite
splits containing $S_k$ since they are automatically
$l$--separable, which reduces the necessary effort to fully
classify a state.

\subsection{Distillability properties within a specific class}

One can also consider the process of distillation (entanglement purification)
and relate it to this classification. Let us consider a specific class,
characterized by all their $k$--separability properties. A necessary condition
for the distillation of a maximally entangled pair e.g. between $A_i$ and $A_j$
is that all those splits for which $A_i$ and $A_j$ belong to different parties
are $k$--inseparable. In fact, it is sufficient to consider only the bipartite splits
fulfilling this property, since this already implies the inseparability of
all $k$--partite splits ($k>2$) of this kind.

In a similar way, we find a necessary condition for the creation of a $j$-GHZ
state, i.e a GHZ state shared among $j$ parties, e.g. $A_{\vec i} \equiv
\{A_{i_0},\ldots,A_{i_j}\}$: We consider all those bipartite splits where not
all of the parties $A_{\vec i}$ are joint at one side. The inseparability of all
those bipartite splits is a necessary condition for the creation of a $j$--GHZ
state between the parties $A_{\vec i}$.

By investigating these necessary conditions for distillability, we
immediately observe that there exists a huge number of classes
which are inseparable at some level (and thus entangled), but
cannot be distilled. Hence all this classes correspond to
different kinds of bound entanglement. For example, we have that
inseparability with respect to any $k$--partite split already
implies that the state is entangled, but one can still have the
2--separability properties such that the necessary conditions for
distillation of a pair between any two parties are not fulfilled.
Sometimes this bound entanglement may be activated by allowing
some additional entanglement between some subsystems. An example
for this is given in Sec.\ref{Example3}.

%-----------------------------------------------------------------------------------
%           Family rho_N
%-----------------------------------------------------------------------------------

\section{Family of states $\rho_N$}\label{family}

In the following we show that any arbitrary $N$--qubit state $\rho$ can be 
brought to a standard form $\rho_N$ \cite{Du99}. We also give a full 
classification of this family of states in terms of their separability and 
distillability properties.

%----------------------------------------------------
\subsection{Notation}
We introduce the orthonormal GHZ-basis \cite{Gr89}
\be
\label{notation}
|\Psi^\pm_j\rangle \equiv \frac{1}{\sqrt{2}} (|j\rangle|0\rangle
  \pm |(2^{N-1}-j-1)\rangle|1\rangle),
\ee
where $j=j_1j_2\ldots j_{N-1}$ is understood in binary notation. We have that
$|j\rangle_{A_1\ldots A_{N-1}}$ is the state of the first $(N-1)$ qubits. For
example, for $N=5$ and $j=6$ this reads $|\Psi^\pm_6\rangle \equiv
\frac{1}{\sqrt{2}} (|0110\rangle_{A_1\ldots A_4}|0\rangle_{A_5} \pm
|1001\rangle_{A_1\ldots A_4}|1\rangle_{A_5})$, since ($6 \equiv 0110$) in binary
notation. Each basis state is a GHZ state, and all basis elements are connected
by $N$--local unitary operations. So
$|\Psi_0^+\rangle=\frac{1}{\sqrt{2}}(|0\ldots0\rangle+|1\ldots1\rangle)$ is only
an arbitrary GHZ state, which can be selected by the choice of a local basis in
$A_1 \ldots A_N$. We emphasis this here, since the states $|\Psi^\pm_0\rangle$
seem to play a special role in what follows. In the following we consider the
family of states
\bea
\label{rhoN}
\rho_N &=& \sum_{\sigma=\pm} \lambda_0^\sigma |\Psi^\sigma_0\rangle\langle
  \Psi^\sigma_0| \nonumber\\
&& + \sum_{j=1}^{2^{(N-1)}-1} \lambda_j (|\Psi^+_j\rangle\langle \Psi^+_j|
  + |\Psi^-_j\rangle\langle \Psi^-_j|),
\eea
which is the straightforward generalization of the family $\rho_3$ introduced in
\cite{Du99} to $N$ qubits. Due to the normalization condition tr($\rho_N$)=1, we
have that $\rho_N$ is described by $2^{N-1}$ independent real parameters. The
labeling is chosen such that $\Delta\equiv \lambda_0^+- \lambda^-_0 \ge
0$.

%----------------------------------------------------
\subsection{Depolarization to $\rho_N$}\label{Depolarization}

In this Section we are going to show that an arbitrary state
$\rho$ can be depolarized to the standard form (\ref{rhoN}) by a
sequence of $N$--local operations while keeping the values of
$\lambda_0^\pm\equiv \langle \Psi^\pm_0|\rho|\Psi^\pm_0\rangle$
and $2\lambda_j \equiv \langle \Psi^+_j|\rho|\Psi^+_j\rangle +
\langle \Psi^-_j|\rho|\Psi^-_j\rangle$ unchanged. Similarly as in
the three--qubit case, this implies that the necessary and
sufficient conditions for distillability and non--separability
found for $\rho_N$ automatically translate into sufficient
conditions for arbitrary states.

We will now explicitly construct the required sequence of
$N$--local operations to obtain the desired depolarization
procedure. By mixing we understand in the following that a certain
operation is (randomly) performed with $p=\frac{1}{2}$, while with
$p=\frac{1}{2}$ no operation is performed. The following $N$
rounds of mixing operations are sufficient to make $\rho$ diagonal
in the basis (\ref{notation}) without changing the diagonal
coefficients: In the first round we apply simultaneous spin flips
at all $N$ locations; The result of this mixing operation is that
all off--diagonal elements of the form
$|\Psi_k^+\rangle\langle\Psi_l^-|$ and
$|\Psi_k^-\rangle\langle\Psi_l^+|$ are eliminated. The remaining
($N-1$) rounds consist of applying $\sigma_z$ to particles $A_k$
and $A_N$ (and the identity to all other particles), where $k$
runs from 1 to $(N-1)$. The effect of the $k^{\em th}$ operation
is the following: a state $|\Psi_j^\pm\rangle$ picks up a minus
sign if $j$, written in binary notation, has a ''1'' at the
$k^{\em th}$ position and remains unchanged if it has a ''0''
there. Since the corresponding $j$ and $i$ of two different basis
states $|\Psi_j^\pm\rangle$ and $|\Psi_i^\pm\rangle$ differ in at
least on digit, this implies that in at least one mixing round one
state, say $|\Psi_j^\pm\rangle$ will pick up a minus sign while
the other state, say $|\Psi_i^\pm\rangle$ will remain unchanged.
This ensures that all off--diagonal elements of the form
$|\Psi_j^\pm\rangle\langle\Psi_i^\pm|$ are eliminated in this
mixing round. Finally, we have that after all $N$ mixing rounds,
$\rho$ is diagonal in the basis (\ref{notation}).

It remains to depolarize the subspaces spanned by $\{|\Psi_j^{\pm}\rangle\}$ for
each $j>0$. This can be accomplished by using random operations that change
$|0\rangle_\alpha\to e^{i\phi_\alpha} |0\rangle_\alpha$
($\alpha=A_1,\ldots,A_N$) with $\sum_k \phi_{A_k}=2\pi$ (this condition ensures
that $\lambda_0^{\pm}$ remains unchanged). This implies that an
arbitrary state $\rho$ can be brought to the standard form $\rho_N$ by a
sequence of $N$--local operations.

One can readily check that the partial transpose of $\rho_N$ with respect to the 
bipartite split $(A_1 \ldots A_{N-1})-A_N$ is positive iff $\Delta\equiv 
\lambda_0^+ -\lambda_0^-\le 2\lambda_{2^{N-1}-1}$. Similar conditions hold for 
each possible bipartite split, i.e. $\rho_N$ has PPT with respect to a certain 
bipartite split iff $\Delta \le 2\lambda_k$ for a specific (unique) $k$ 
corresponding to this bipartite split. To determine the corresponding $k$, let 
us consider a bipartite split where $l$ qubits $A_{\vec k} \equiv\{A_{k_1}, 
\ldots A_{k_l}\}$ are jointly located at one side, while the remaining $N-l$ 
qubits are located at the other side. Without loss of generality we can assume 
that $A_N \notin A_{\vec k}$. In this case, the corresponding $\lambda_k$ is 
given by $k$, which, written in binary notation, has ones at the positions 
($k_1, \ldots ,k_l$) and zeros at all other positions and only the highest 
$(N-1)$ bits are considered (the lowest one, $k_N$, is allways zero because we 
assumed that $A_N \notin A_{\vec k}$). For instance we consider $N=6$ and the 
bipartite split $S_2=(A_1A_4A_5)-(A_2A_3A_6)$. We have that the corresponding $k$ 
is given by $k=100110$, where we only have to consider the highest 5 bits. Thus 
$k=10011 \equiv 19$ and we have that the state $\rho_N$ has PPT with respect to the 
bipartite split $S_2$ iff $\Delta \le 2\lambda_{19}$

%----------------------------------------------------
\subsection{Separability of $\rho_N$}\label{Separability}

Let us enumerate the separability properties of the states (\ref{rhoN}). Then we obtain
\\
(i) We consider a specific $k$--partite split $S_k$ of $\rho_N$. Iff all
bipartite splits which contain $S_k$ have PPT, then $\rho_N$ is $k$--separable with
respect to this specific $k$--partite split.

In order to prove (i), we find it useful to consider first two special cases of 
(i) in order to illustrate the underlying ideas. These special cases are the 
following:
\\
(ii) We consider a specific bipartite split, where we have that $l$ qubits $A_{\vec k}
\equiv\{A_{k_1}, \ldots A_{k_l}\}$ are jointly located at one side, while the
remaining $N-l$ qubits are located at the other side. Iff we have that the
partial transpose corresponding to this bipartite split is positive, i.e.
$\rho_N^{T_{A_{\vec k}}} \equiv \rho_N^{T_{A_{k_1}} \ldots T_{A_{k_l}}}  \ge 0$
then $\rho_N$ is fully separable with respect to this bipartite split, i.e. it  can be
written in the form
\be
\label{form}
\rho_N = \sum_i |\chi_i\rangle_{A_{\vec k}}\langle \chi_i|
  \otimes |\varphi_i\rangle_{\rm rest}\langle \varphi_i|.
\ee
(iii) We consider all possible $2^{N-1}-1$ bipartite splits of a $N$--qubit
system. Iff for each of those splits the corresponding partial transposition is
positive, then $\rho_N$ is fully separable, i.e. $\rho_N$ is $N$--separable.

These statements are illustrated for the simplest cases of a $3$ and $4$--qubit system in
Sec. \ref{Examples}. From (i) follows that $k$--separability with respect to a certain $k$--partite 
split $S_k$ of the states $\rho_N$ is implied by the 2--separability properties 
of the bipartite splits containing $S_k$. Thus the family $\rho_N$ is completely 
characterized by its 2--separability properties, which already determine the 
hierarchic structure proposed in \ref{Classification}.

In the remainder of this section, we are going to proof the
statements (i)-(iii). Let us start by proving (ii). The basic idea of the proof
is to define a state $\hat\rho$ which we show to be 2--separable with respect to
the bipartite split in question and which can be depolarized to $\rho_N$. Since
a separable state is converted into a separable one by depolarization (which is
a $N$--local process), this automatically implies the 2--separability of
$\rho_N$. We have that $\rho_N$ has positive partial transposition with respect
to the bipartite split $A_{\vec k}$ iff $\Delta \le 2\lambda_k$ for $k$
corresponding to $\vec k$.
We define
\be
\hat\rho= \rho_N + \frac{\Delta}{2} (|\Psi^+_k\rangle\langle \Psi^+_k|-
|\Psi^-_k\rangle\langle \Psi^-_k|).
\ee
The state $\hat \rho$ is positive since $\Delta \le 2\lambda_k$ and has the
property $\hat\rho = \hat\rho^{T_{A_{\vec k}}}$. For a bipartite split of the form (one
qubit)-rest, this already implies the separability of $\hat \rho$, since it has
been shown in \cite{Le99} that all states in $\C^2 \otimes \C^N$ which fulfill
$\tilde\rho^{T_A}=\tilde\rho$ are separable. For all the other splits we show the
separability directly. We rewrite $\hat\rho$ as follows
\bea
\hat\rho&=&\Delta(|\Psi^+_0\rangle\langle \Psi^+_0|+|\Psi^+_k\rangle\langle \Psi^+_k|) \nonumber \\
&+&(\lambda_k-\frac{\Delta}{2})(|\Psi^-_k\rangle\langle \Psi^-_k|+|\Psi^+_k\rangle\langle \Psi^+_k|) \label{li} \nonumber \\
&+&\lambda_0^-(|\Psi^-_0\rangle\langle \Psi^-_0|+|\Psi^+_0\rangle\langle \Psi^+_0|) \nonumber\\
&+& \sum_{j=1, j\not=k}^{2^{(N-1)}-1} \lambda_j (|\Psi^+_j\rangle\langle \Psi^+_j|
  + |\Psi^-_j\rangle\langle \Psi^-_j|),
\eea
We have that all prefactors are positive (since $\Delta \le 2\lambda_k$). The
terms in lines 2-4 are completely separable, which can be seen by using
that $(|\Psi^+_j\rangle\langle \Psi^+_j| + |\Psi^-_j\rangle\langle
\Psi^-_j|)=|j0\rangle\langle j0|+|(2^{N-1}-j-1)1\rangle\langle(2^{N-1}-j-1)1|$.
The term in line 1 is biseparable with respect to the bipartite split $\vec k$.
To see this, let us rewrite the basis states as follows:
\bea
|\Psi_0^+\rangle &=&\frac{1}{\sqrt{2}} (|0\ldots0\rangle_{A_{\vec k}}|0\ldots0\rangle_{\rm rest}+|1\ldots1\rangle_{A_{\vec k}}|1\ldots1\rangle_{\rm rest}) \nonumber \\
|\Psi_k^+\rangle &=&\frac{1}{\sqrt{2}} (|1\ldots1\rangle_{A_{\vec k}}|0\ldots0\rangle_{\rm rest}+|0\ldots0\rangle_{A_{\vec k}}|1\ldots1\rangle_{\rm rest}).
\eea
We define $|\vec \pm\rangle = \frac{1}{\sqrt{2}}(|0\ldots 0\rangle \pm |1\ldots
1\rangle)$. It is now straightforward to check that line one of (\ref{li}) can be written as
$
\Delta(|\vec +\rangle_{A_{\vec k}}\langle\vec +| \otimes |\vec +\rangle_{\rm rest}\langle \vec +|+
|\vec -\rangle_{A_{\vec k}}\langle\vec -| \otimes |\vec -\rangle_{\rm rest}\langle \vec -|),
$
which is clearly biseparable with respect to the bipartite split $\vec k$ and
concludes the proof in one direction. If we consider on the other hand that
$\rho_N$ is biseparable with respect to the bipartite split $\vec k$, it follows
trivially that it also has PPT corresponding to this bipartite split, since
positive partial transposition is a necessary condition for
separability\cite{Pe96}.

To prove the third statement (iii), we show that if $\rho_N$ has PPT with respect
to all possible bipartite splits then $\rho_N$ is $N$--separable (note again that
the opposite is trivially true). This condition implies that $\Delta/2\le
\lambda_j$ for all $j$.  Again, the idea is to define an
operator $\tilde\rho$ which can be depolarized into the form
$\rho_N$ by using local operations and that is fully separable.
Let $\tilde\rho$ be a state of the form (\ref{rhoN}) with
coefficients $\tilde
\lambda_0^\pm \equiv \lambda_0^\pm$, and $\tilde\lambda_k^\pm \equiv \lambda_k \pm
\Delta/2$ ($k=1,\ldots,(2^{N-1}-1)$). Clearly, $\tilde\rho$ can be depolarized into $\rho_N$.
We now rewrite $\tilde\rho$ as follows:
\bea
\tilde\rho&=&\frac{1}{2}\sum_{k=0}^{2^{N-1}-1}
(\tilde\lambda^+_k+\tilde\lambda^-_k-\Delta)(|\Psi_k^+\rangle\langle \Psi_k^+| +
  |\Psi_k^-\rangle\langle \Psi_k^-|) \nonumber \\
 &+& \Delta \sum_{k=0}^{2^{N-1}-1} |\Psi_k^+\rangle\langle\Psi_k^+|.\label{sepa}
\eea
Since all possible partial transposes are positive, we have that all
coefficients in (\ref{sepa}) are positive. The first term in (\ref{sepa}) can be
written as $\sum_{k=0}^{2^N-1}
\frac{(\tilde\lambda^+_k+\tilde\lambda^-_k-\Delta)}{2} (|k,0\rangle\langle k,0|+
|(2^{N-1}-k-1),0\rangle\langle (2^{N-1}-k-1),0|)$ and is thus fully separable.
It remains to show that the second term in (\ref{sepa}) is also $N$--separable.
Let us first define the states $|\pm\rangle=(|0\rangle\pm|1\rangle)/\sqrt{2}$.
To show the separability of the second term, we write it as $\sum_{j=0}^{2^N-1}
|\phi_j\rangle\langle\phi_j|$ where $|\phi_j\rangle$ are all the states of the
form $|\sigma_1\sigma_2,\ldots\sigma_N\rangle$ with $\sigma_i=\pm$, which have
an even number of minuses. All the states $|\phi_j\rangle$ are fully
separable, which concludes the proof.

Now we are ready to prove the first statement (i). The basic idea of the
proof is very similar to the one used in the previous proofs: We define a state
$\rho'$ which can be depolarized to $\rho_N$ and we show that
$\rho'$ is $k$--separable. We have that a number of bipartite
splits have PPT. To a specific bipartite splits corresponds the
relation $\Delta \le 2\lambda_{i_0}$, which is the condition that
this specific bipartite split has PPT. Thus we have that $\Delta
\le 2\lambda_{i}$, where $i \in \{i_0,\ldots,i_l\} \equiv
\vec{i}$, and each $\lambda_i$ corresponds to a bipartite split
which does not further divide systems which were joint for the
$k$--partite split we consider.   Let $\rho'$ be a state of the
form (\ref{rhoN}) with coefficients $\lambda_0^{' \pm}  \equiv
\lambda_0^\pm$, $\lambda_i^{' \pm}  \equiv \lambda_i \pm \Delta/2$
for $i \in \vec{i}$ and $\lambda_k^{' \pm} \equiv \lambda_k$ for
$k \notin\vec{i}$. Clearly, $\rho'$ can be depolarized to
$\rho_N$. Similarly as in the previous proofs, we rewrite $\rho'$
as follows: \bea \rho'&=& \Delta(|\Psi^+_0\rangle\langle
\Psi^+_0|+\sum_{i \in \vec i}|\Psi^+_i\rangle\langle \Psi^+_i|)
\nonumber \\ &+&\sum_{i \in \vec
i}(\lambda_i-\frac{\Delta}{2})(|\Psi^-_i\rangle\langle
\Psi^-_i|+|\Psi^+_i\rangle\langle \Psi^+_i|) \label{li1} \\
&+&\sum_{k\notin \vec i} \lambda_k (|\Psi^+_k\rangle\langle
\Psi^+_k|
  + |\Psi^-_k\rangle\langle \Psi^-_k|) \nonumber \\
&+&\lambda_0^-(|\Psi^-_0\rangle\langle \Psi^-_0|+|\Psi^+_0\rangle\langle \Psi^+_0|) \nonumber.
\eea
All coefficients in (\ref{li1}) are positive, and lines (2-4) are fully
separable. It remains to show the $k$--separability of line 1 of (\ref{li1}). To
see this, we define the states $|\vec \pm\rangle_{{\rm party} l} =
\frac{1}{\sqrt{2}}(|0\ldots 0\rangle_{{\rm party} l} \pm |1\ldots 1\rangle_{{\rm
party} l})$, and the states $|\Phi_j\rangle \equiv |\sigma_1\rangle_{{\rm party}
1} \otimes \ldots \otimes|\sigma_k\rangle_{{\rm party} k}$ with $\sigma_i=\pm$
and the number of minuses is even. It is now straightforward to check that line
1 of (\ref{li1}) can be written as $\Delta\sum_j |\Phi_j\rangle\langle\Phi_j|$
and is thus $k$--separable with respect to the $k$--partite split we consider,
and concludes the proof in one direction.

If we consider on the other hand that $\rho_N$ is $k$--separable with respect to 
a specific $k$--partite split $S_k$, it follows that it is also biseparable with 
respect to all bipartite splits which contain $S_k$, since any of those 
bipartite splits corresponds to joining systems which were divided for the 
$k$--partite split. But the positivity of the partial transposition is a 
necessary condition for biseparability corresponding to a certain bipartite 
split\cite{Ho97}, from which follows that positivity of all bipartite splits we 
consider is also a necessary condition for $k$--separability. So again the 
conditions we found are necessary and sufficient.

%----------------------------------------------------
\subsection{Distillability of $\rho_N$}\label{Distillability}

We will turn now to analyze the distillability properties of $\rho_N$:
\\
(i) We consider all possible bipartite splits of the $N$ qubits where the
particles $A_i$ and $A_k$ belong to different parties. Iff all such splits have
negative partial transposition, then a maximally entangled pair between particle
$A_i$ and $A_k$ can be distilled.
\\
(ii) We consider the $k$ parties $A_{\vec i} \equiv \{A_{i_0},\ldots,A_{i_k}\}$
($k \leq N$) and consider all those bipartite splits where not all of the
parties $A_{\vec i}$ are joint at one side. Iff all those splits have negative
partial transposition, then a $k$--GHZ state (i.e. a GHZ state shared between
$k$ parties) between the parties $A_{\vec i}$ can be distilled.

To show (i) and without loss of generality we take $i=N$ and $k=N-1$. In that
case, the condition we impose on the partial transpositions is equivalent to
require that $\Delta/2 > \lambda_j$ with $j$ odd [note that with the notation we
are using, the state of the $N-1$ qubit determines the parity of the states
$|j\rangle$ in (\ref{notation})]. In order to show the distillability of a
maximally entangled state between $A_{N-1}$ and $A_N$, it is sufficient to show
that a pair with fidelity $F>0.5$ (overlap with the maximally entangled state
$|\Phi^+\rangle$) between those two parties can be created \cite{Be96}. If we
project all the qubits except the ones at $A_N$ and $A_{N-1}$ onto the state
$|+\rangle$ we see that the resulting state obtained from $\rho_N$ has $F>0.5$
and can thus be distilled to a maximally entangled state between $A_N$ and
$A_{N-1}$ iff
\be
\Delta/2 > \sum_{j {\rm odd}}\lambda_j\label{distillcondition}.
\ee
Eventhough we have that $\Delta/2 > \lambda_j$ for all $j$ odd, the condition
(\ref{distillcondition}) might not be fulfilled. In this case we use the
following purification procedure: The idea is to combine $M$ systems in the same
state $\rho_N$, perform a measurement and obtain one system with the same form
(\ref{rhoN}) but in which the new $\Delta$ is exponentially amplified with
respect to $\lambda_{k}$, $k$ odd. In order to do that, let us define the
operator
\be
P=|00\ldots 00\rangle\langle 00\ldots 00| + |10\ldots 00\rangle\langle 11\ldots 11|,
\ee
which acts on $M$ qubits. Now we proceed as follows: We take $M$ systems, and 
apply the operator P in all $N$ locations. This corresponds to measuring a POVM 
that contains $P$ obtaining the outcome associated to $P$. The resulting state 
$P^{\otimes N}\rho_N^{\otimes M}(P^\dagger)^{\otimes N}$ has the first system in 
an (unnormalized) density operator of the form (\ref{rhoN}) but with new 
coefficients $\tilde\Delta$ and $\tilde\lambda_k$. In order to calculate these 
new coefficients, we need the following observations: First, the operator 
$\rho_N^{\otimes M}$ is diagonal in the basis $\{|\chi_{k_0 \ldots 
k_M}^{\sigma_1 \ldots \sigma_M}\rangle\}$ with coefficients 
$\lambda_{k_1}^{\sigma_1}\ldots \lambda_{k_M}^{\sigma_M}$ where
\be
|\chi_{k_0 \ldots k_M}^{\sigma_1 \ldots \sigma_M}\rangle =  |\Psi_{k_0}^{\sigma_0}\rangle \otimes \ldots \otimes |\Psi_{k_M}^{\sigma_M}\rangle ,\label{baseM}
\ee
and $k_j\in\{0, \ldots 2^{N-1}-1\}, \sigma_j=\pm$. Second, we need the action
of the operator $P^{\otimes N}$ on the basis
states (\ref{baseM}). We find
\be
P^{\otimes N} |\chi_{k_0 \ldots k_M}^{\sigma_1 \ldots \sigma_M}\rangle = \delta_{k_0 \ldots k_M} |\Psi_{k_0}^{\sigma}\rangle_{\rm system 1}|0 \ldots 0\rangle_{\rm rest},
\ee
where $\sigma=+$ if the number of minuses in $\{\sigma_1 \ldots \sigma_M\}$ is
even, and $\sigma=-$ otherwise. Note also that we only have a contribution if
$k_0=k_1= \ldots =k_M$. Using this results, it is now straightforward to check
that the first system of $P^{\otimes N}\rho_N^{\otimes M}(P^\dagger)^{\otimes
N}$ is an (unnormalized) density operator of the form (\ref{rhoN}) with new
coefficients
\bea
\tilde\Delta/2 = (\Delta/2)^M &;& \tilde\lambda_k=\lambda_k^M.
\eea
Given that $\Delta/2 >\lambda_k$, $k$ odd, for sufficiently large $M$ we have
that condition (\ref{distillcondition}) is fulfilled, i.e. that after the
projection of all systems except $A_{N-1}$ and $A_N$ on the state $|+\rangle$,
the resulting state has $F>0.5$ and is thus distillable, which concludes the
proof in one direction.
On the other hand, the condition we impose for distillability is also necessary.
Having a maximally entangled pair between the parties $A_{N-1}$ and $A_N$
implies that all bipartite splits in question have NPT. Since local operations
keep the positivity of the partial transposition \cite{Ho97b}, we must start
with NPT of all bipartite splits in question.

To prove (ii), we just have to recognize that the condition we
impose guarantees that maximally entangled pairs between any two
parties within $A_{\vec i}$ can be distilled. This is clearly
sufficient to create a GHZ state among those parties, e.g. by
means of teleportation (creating the GHZ state locally at
$A_{i_0}$ and teleporting the $(k-1)$ qubits to the parties
$\{A_{i_1},\ldots,A_{i_k}\}$ using maximally entangled pairs
created among $A_{i_0}$ and $A_{j}$ with $j \in \{i_1,\ldots
i_k\})$. Note also the the condition we impose is also necessary,
since a GHZ state allows us to create maximally entangled pairs
among all parties involved, which implies that all bipartite
splits in question have NPT and thus have to have NPT at the
beginning, since one cannot convert a state from PPT to NPT by
means of local operations \cite{Ho97b}.

\section{Examples}\label{Examples}

%--------------------------------------------------------------------------------
%   Example:        3 qubits
%--------------------------------------------------------------------------------

\subsection{Three--qubit systems}\label{Example3}

In this section we investigate the simplest case of three qubits, each hold by
one of the parties $A,B$ or $C$.

\subsubsection{Classification}
In order to perform the classification proposed in
\ref{Classification}, we consider the 3--partite split of the
system as well as all possible bipartite splits, where all
together three such splits exist. Each corresponds to having one
system (e.g. $A$) on one side and the two other systems (e.g. $B$
and $C$) on the other side. In other words, for this specific
bipartite split of our three-qubit system we allow the parties $B$
and $C$ to act together, i.e. ${\cal
H}=\C^{2}_A\otimes\C^{4}_{BC}$. Thus each of this bipartite splits
will give us an upper limit of what can be done by three--local
operations on the system. To perform the classification, we have
to consider the separability properties of the 3--partite and
bipartite splits. In particular, whether they can be written in
one or more of the following forms: \bma \label{all} \bea
\label{ABC} \rho &=& \sum_i |a_i\rangle_A\langle a_i| \otimes
|b_i\rangle_B\langle b_i|
 \otimes |c_i\rangle_C\langle c_i|\\
\label{A(BC)}
\rho &=& \sum_i |a_i\rangle_A\langle a_i| \otimes |\varphi_i\rangle_{BC}\langle \varphi_i|\\
\label{B(AC)}
\rho &=& \sum_i |b_i\rangle_B\langle b_i| \otimes |\varphi_i\rangle_{AC}\langle \varphi_i|\\
\label{C(AB)}
\rho &=& \sum_i |c_i\rangle_C\langle c_i| \otimes |\varphi_i\rangle_{AB}\langle \varphi_i|.
\eea
\ema
Here, $|a_i\rangle$, $|b_i\rangle$ and $|c_i\rangle$ are (unnormalized) states
of systems $A$, $B$ and $C$, respectively, and $|\varphi_i\rangle$ are states of
two systems. We call a state biseparable with respect to a certain bipartite
split if it is separable with respect to this split, e.g. a state is biseparable
with respect to the bipartite split $A-(BC)$ iff it can be written in the form (\ref{A(BC)}).
Similary, a state is called triseparable (3--separable) if it is separable with
respect to the split $A-B-C$, i.e. can be written in the form (\ref{ABC}).

At the top level of the classification (level 3), we consider the tripartite
split $A-B-C$ and determine whether the state $\rho$ is 3--separable or not. At
the second level (level 2), one considers all possible bipartite splits $[A-(BC), B-(AC), C-(AB)]$ and
determines whether the state $\rho$ can be written in one or more of the forms
(\ref{A(BC)},\ref{B(AC)},\ref{C(AB)}). At this level of the classification, one has $2^3=8$
different possibilities, each corresponding to a physically different situation.
For arbitrary three--qubit systes one thus finds the following complete set of 9
disjoint classes

{\bf Class 1} {\it Fully inseparable states}: states that cannot be written in
any of the above forms (\ref{all}). An example is the GHZ state \cite{Gr89}
$|\Psi^+_0\rangle$.

{\bf Classes 2.1, 2.2, 2.3} {\em 1-qubit biseparable states:}
Class 2.1: biseparable states with respect to qubit $A$ are states
that are separable in $A-(BC)$, but non--separable otherwise. That
is, states that can be written in the form (\ref{A(BC)}) but not
as (\ref{B(AC)}) or (\ref{C(AB)}). An example is the state
$|0\rangle_A\otimes|\Phi^+\rangle_{BC}$, where
$|\Phi^+\rangle=(|00\rangle+|11\rangle)/\sqrt{2}$ is a maximally
entangled state of two qubits. Similarly, class 2.2 and 2.3
correspond to biseparable states with respect to qubit $B$ and
$C$ respectively.

{\bf Classes 3.1, 3.2, 3.3} {\it 2-qubit biseparable states}:
Class 3.1: biseparable states with respect
to qubits $A$ and $B$ are states that are separable in $A-(BC)$ and $B-(AC)$, but
non--separable in $C-(AB)$. That is, states that can be written in the forms
(\ref{A(BC)}) and (\ref{B(AC)}) but not as (\ref{C(AB)}). For examples, see
below. Similary, classes 3.2 (3.3) are biseparable with respect to the qubits $A$ and $C$ ($B$ and $C$).

{\bf Class 4} {\it 3-qubit biseparable states}:
Those are states that are separable in $A-(BC)$, $B-(AC)$ and $C-(AB)$ (i.e. separable
with respect to each bipartite split), but which are not completely separable,
i.e. cannot be written as (\ref{ABC}). For an example, see Ref.\ \cite{Be98}.

{\bf Class 5} {\em Fully separable states}:
These are states that can be written in the form (\ref{ABC}) and are thus also
separable with respect to each bipartite split. A trivial example is a product
state $|1\rangle_A\otimes|1\rangle_B\otimes|1\rangle_C$.

Note that the classes {\bf 2.1, 2.2, 2.3} (respectively {\bf 3.1, 3.2, 3.3}) where
identified in \cite{Du99}, since they can be obtained from each other by
permuting the parties. In this case, 5 distinct classes remain.

\subsubsection{Family $\rho_3$}

Let us now concentrate on the family of three--qubit states $\rho_3$
(\ref{rhoN}). This family is characterized by 4 parameters,
$\{\Delta\equiv\lambda_0^+ - \lambda_0^-,\lambda_1,\lambda_2,\lambda_3\}$, and we
have that any state can be depolarized to this standard form.
The GHZ basis (\ref{notation}) reads in this case
\be
|\Psi^\pm_j\rangle \equiv \frac{1}{\sqrt{2}} (|j\rangle_{AB}|0\rangle_C
  \pm |(3-j)\rangle_{AB}|1\rangle_C), \label{GHZbasis}
\ee where $|j\rangle_{AB}\equiv |j_1\rangle_A|j_2\rangle_B$ with
$j=j_1j_2$ in binary notation. For example,
$|\Psi_0^{\pm}\rangle=\frac{1}{\sqrt{2}}(|000\rangle \pm
|111\rangle)$ are standard GHZ states, as well as
$|\Psi_3^{\pm}\rangle=\frac{1}{\sqrt{2}}(|110\rangle \pm
|001\rangle)$ ($3 \equiv 11$ in binary notation). Note that all 8
basis states are connected by 3--local unitary operations, i.e.
each basis state is a maximally entangled GHZ state. Due to the
fact that the local bases in $A, B$ and $C$ can be chosen
arbitrarily, none of the basis states has any preferences.

We will now investigate the separability and distillability
properties of $\rho_3$ and give a full classification in terms of
the classes introduced above. It turns out that using the partial
transpose criterion for each bipartite split characterizes the
state $\rho_3$ completely. The conditions under which the operator
$\rho_3$ has positive partial transpose with respect to each qubit
are as follows \bea \label{pt} \rho_3^{T_A} \ge 0 \quad &{\rm
iff}& \ \Delta \le 2 \lambda_{2} \nonumber \\ \rho_3^{T_B} \ge 0
\quad &{\rm iff}& \ \Delta \le 2 \lambda_{1} \\ \rho_3^{T_C} \ge 0
\quad &{\rm iff}& \ \Delta \le 2 \lambda_{3}.\nonumber \eea Recall
that each of this conditions correspond to a (virtually) bipartite
split of the system. Investigating e.g. $\rho_3^{T_A}$, we
actually have in mind a bipartite split of the system into $A$ on
one side and $BC$ on the other side. From these conditions we also
see that in general no further depolarization which keeps the form
of $\rho_3$ is possible (except the depolarization toward the
completely depolarized state, which can always be done trivially).
We show this by giving a counterexample. Imagine we would like to
depolarize the subspaces spanned by
$\{|\Psi_1^\pm\rangle,|\Psi_2^\pm\rangle\}$ and thereby equalize
the coefficients $\lambda_1$ and $\lambda_2$. For certain values
of the parameters, this would imply that the state after this
depolarization has negative partial transposition with respect to
one party, while it started with positive partial transposition.
Since one cannot change the positivity of the partial transpose by
local operations\cite{Ho99}, this further depolarization is
impossible in general. (e.g. $\rho_3$ with
$\lambda_0^+=\frac{2}{3},\lambda_2=\frac{1}{3}$, all other
parameter 0 has $\rho_3^{T_A} \geq 0$, but would have negative
partial transposition with respect to all bipartite splits after
the (imaginary) further depolarization in question)

%----------------------------------------------------
\subsubsection{Separability of $\rho_3$}\label{Sep3}
Let us specialize the theorems about separability obtained in \ref{Separability} 
to $N=3$. In this case, we do not need the general theorem (i), but only 
give examples for the statements (ii) and (iii):
\\
(ii) $\rho_3$ is separable with respect to the bipartite split $A-(BC)$, i.e. it can be
written in the form (\ref{A(BC)}) iff $\rho_3^{T_A}\ge 0$ [and analogously for
(\ref{B(AC)}) and (\ref{C(AB)}) with $\rho_3^{T_B}\ge 0$ and $\rho_3^{T_C}\ge
0$, respectively].
\\
(iii) $\rho_3$ is completely separable, i.e. it can be written as
(\ref{ABC}) iff $\rho_3^{T_A},\rho_3^{T_B},\rho_3^{T_C} \ge 0$.
\\
Note that these are {\it iff} statements and thus provide a full
characterization of $\rho_3$ in terms of the separability properties. The
resulting classification is summarized in Table \ref{Table1}. Here we have that
3--qubit biseparabilty implies fully separabilty (tri--separability), while in
general this is not necessarily the case (otherwise class 4 would be empty). We
have the the family $\rho_3$ is completely characterized by its biseparability
properties, so only the structure at level 2 is necessary for classification.

Furthermore, this also provides us with sufficient conditions for
non--separabilty for arbitrary states $\rho$. Namely if there exist a basis
(\ref{GHZbasis}) such that the corresponding state $\rho_3$ after depolarization
has e.g. the property that $\rho_3^{T_A}$ is negative, this implies that $\rho$
is non--separable in $A-(BC)$. Note also that no conclusion can be drawn about the
separability properties of $\rho$ given PPT of the depolarized state $\rho_3$, since the
depolarization process might convert a non--separable state $\rho$ into a
separable state $\rho_3$.

%----------------------------------------------------
\subsubsection{Distillability of $\rho_3$}
We now state the distillability properties of $\rho_3$:
\\
(i) One can distill a maximally entangled state $|\Phi^+\rangle_{\alpha\beta}$ between $\alpha$ and $\beta$ iff both
$\rho_3^{T_\alpha},\rho_3^{T_\beta}$ have negative partial transposition.
\\
(ii) Iff all three partial transposes are negative, we can distill
a GHZ state (since we can distill an entangled state between $A$
and $B$ and another between $A$ and $C$ and then connect them to
produce a GHZ state \cite{Zu93}).
\\
(iii) If we have that $\rho_3^{T_C}$ is negative but $\rho_3^{T_A},\rho_3^{T_B}\ge
0$ (i.e. have PPT) and we have maximally entangled states between $A$ and $B$ at our
disposal, then we can {\it activate} the entanglement between $ABC$ and create
a GHZ state.

Note that (iii) is an example for the activation of bound entanglement. In this
example, the state is inseparable with respect to the bipartite split $C-(AB)$
and thus entangled. However no entanglement between any two subsystems can be
created, since we have that the PPT of $A$ and $B$ implies the separability of
$A-(BC)$ and $B-(AC)$ (see (ii) in Sec. \ref{Sep3}). However entanglement between
$A$ and $B$ is sufficient to allow the creation of a GHZ state.
One can show this by noting that the singlets allow to teleport states between
locations $A$ and $B$. Thus, for all practical purposes we can consider a pair
of qubits $AB$ as a four--level system in which we can perform arbitrary
operations. The situation is equivalent to that in which one has 2--level
systems entangled to 4--level systems such that the density operator describing
one pair acts on $\C^2\otimes \C^4$ and has a negative partial transpose. It can
be easily shown \cite{Du99b} that in systems $2\times N$ negative partial
transpose is a necessary and sufficient condition for distillation, and
one can thus distill arbitrary states. Using again teleportation, one can
end up with a GHZ state shared by $A$, $B$ and $C$.

Again, these results also provide us with sufficient conditions for
distillability of arbitrary states $\rho$. From (i) follows: If there exist a
basis (\ref{GHZbasis}) such that the corresponding state $\rho_3$ after
depolarization has e.g. the property that $\rho_3^{T_A},\rho_3^{T_B}$ are
negative, this implies that a maximally entangled pair between $A$ and $B$ can be
distilled from $\rho$. Note here that to make this condition for distillability
necessary {\it and} sufficient for arbitrary states $\rho$, one should find a
depolarization procedure (may be assisted by some appropriate local filtering
operations) such that the NPT property is maintained. That is, the corresponding
state $\rho_3$ after depolarization should still have
$\rho_3^{T_A},\rho_3^{T_B}$ negative.

Note also that a sufficient condition to distill a GHZ-state from
$\rho$ is that there exist two different bases (\ref{GHZbasis})
such the corresponding states $\rho_3$ after depolarization only
have two of the partial transposes kept negative, where for the
second basis one has to differ from the first one. For example,
basis 1 allows us to keep $A$ and $B$ each having NPT (and thus to
distill a pair between $A$ and $B$), while basis 2 keeps $B$ and
$C$ having NPT. This together ensures that a GHZ state can be
created. Clearly, one has to start with $\rho$ which has all three
partial transposes negative. Again, one may try to make this
condition necessary and sufficient by introducing a (filtering
assisted) depolarization procedure such that at least two of the
partial transposes are kept negative for a certain choice of
basis.

%------------------------------------------------------------------------------------------
%------------------------------------------------------------------------------------------
%  Example: 4 qubit system
%------------------------------------------------------------------------------------------
%------------------------------------------------------------------------------------------

\subsection{Four--qubit system}\label{Example4}

Here we consider the special case of a 4--partite system in order to illustrate the
theorems about separability and distillability obtained in the previous
sections. For convenience, let us call the parties $A,B,C$ and $D$ instead of $A_1,
\ldots ,A_4$.

\subsubsection{Classification}
We start by illustrating the classification for general 4--qubit systems. At the
top level of the structure is the 4--separability, that is the question whether
the state $\rho$ is separable with respect to the 4--partite split $(4)
A-B-C-D$, i.e. fully separable. At the second level, we have to consider 6
different 3--partite splits of the system: (3a) $A-B-(CD)$, (3b) $A-(BC)-D$,
(3c) $A-(BD)-C$, (3d) $(AB)-C-D$, (3e) $(AC)-B-D$ and (3f) $(AD)-B-C$. At the
third level, all together 7 different 2--partite splits exist, namely (2a)
$A-(BCD)$, (2b) $B-(ACD)$, (2c) $C-(ABD)$, (2d) $D-(ABC)$, (2e) $(AB)-(CD)$, (2f)
$(AC)-(BD)$ and (2g) $(AD)-(BC)$.

We have for instance that the 3--partite split (3a) is contained in the
2--partite splits (2a), (2b) and (2e). All other bipartite splits
cannot be otained from (3a) by joining some of the parties, since
they would divide the system $(CD)$ along different parties, and
thus the tripartite split (3a) does not belong to them.

The classification thus takes place as follows (see also Fig. \ref{Fig1}): At 
the top level (level $4$), one has to decide whether the state is 4--separable 
or not. In the case it is 4--separable, it automatically follows that it is also 
3-- and 2--separable with respect to all possible 3-- or 2--partite splits. If 
it is 4--inseparable, one has to investigate the various kinds of 
3--separability at level 3, where one can have all possible combinations of the 
6 kinds (3a)-(3f) of 3--separability and and 3--inseparability. We have $2^6$ 
different configurations at this level. At the next level of the classification 
(level 2), one investigates all possible bipartite splits (2a)-(2g) closer. One 
finds $2^7$ different configurations at this level.

The structure at level 2 is partly determined by the
structure at level 3 and vice versa. If, e.g., the state is
3--separable with respect to the 3--partite split (3a), it follows
that the bipartite splits (2a),(2b) and (2e) at level 2 which
contain (3a) are also 2--separable. In the case where a state is
3--separable with respect to the splits (3a), (3b) and (3c), it
even follows that the state is 2--separable with respect to all
possible bipartite splits. Although it still can be 3--inseparable
with respect to the 3--partite spits (3d),(3e) or (3f) in
principle, the underlying structure at level 2 is already
completely determined by the structure at level 3.
3--inseparability with respect to a specific 3--partite split
$S_3$, on the other hand, still allows all combinations of
2--separability and 2--inseparability within the bipartite splits
at level 2 which contain $S_3$. Conversely, 2-inseparabilty with
respect to the bipartite split (2a) implies 3--inseparability with
respect to the 3--partite splits (3a), (3b) and (3c), while
2--separability still leaves all possibilities at level 3 open.
From this one also sees that it is neither sufficient to consider
only the $4$-- and $2$--separability to classify the state
completely, nor to consider only 4-- and 3--separability.

Furthermore, it is not sufficient to classify the states by the number of
$k$--separable state at level $k$ of the hierarchic classification. For example,
we have that 3--separability with respect to the 3--partite splits (3a),(3b) and
(3c) already implies 2--separability with respect to all bipartite splits. On
the other hand, 3--separability with respect to the 3--partite splits (3a),(3b)
and (3d) does not determine the biseparability properties of the bipartite split
(2f), which may thus still be 2--inseparable. The two kinds of 3 times
3--separability correspond to different physical situations, and one cannot
obtain one configuration from the other one by permuting the parties. Thus it is
not sufficient to give only the number of 3--separable states, one also needs
the information which of the splits is separable and which is not.

\subsubsection{Separability and distillability properties of $\rho_4$}
Let us now turn to the family of states $\rho_4$ and illustrate its separability
properties. We will give an example for each theorem.
\\
(i) Let us consider the 3--partite split $A-B-(CD)$ (3a). Iff we have that the
2--partite splits (2a), (2b) and (2e) (which contain (3a)) have PPT, then
$\rho_4$ is 3--separable with respect to this 3--partite split.
\\
(ii) Iff the partial transposition with respect to the bipartite split
$(AB)-(CD)$ (2e) is positive, then $\rho_4$ is 2--separable in $(AB)-(CD)$.
\\
(iii) Iff for all possible
2--partite splits (2a)-(2g) we have that the corresponding partial transposition
is positive, then $\rho_4$ is 4--separable.

Note that this family of states is completely characterized by its
2--separability properties, since from 2--separability follows the
corresponding 3--separability as well as the corresponding
4--separability. So in this case only one level of the hierarchic
structure, namely level 2, is required to fully classify the
states $\rho_4$.

Finally we consider the distillability properties of $\rho_4$.
\\
(i) Iff the partial transposition with respect to the bipartite splits
(2c), (2d), (2f) and (2g) is negative, then a maximally entangled pair between
$C$ and $D$ can be distilled. Note that these four splits are the only ones of
relevance, since the parties $C$ and $D$ are not joint there.
\\
(ii) Iff the partial transpositions with respect to the bipartite splits (2a),
(2b), (2c), (2e), (2f) and (2g) are negative, then a GHZ state between the
parties $A-B-C$ can be distilled. Note that the split (2d) is the only one which
is not of relevance here, since the parties $ABC$ are joint in this case. If in
addition also the split (2d) has NPT, then a GHZ state between all four parties
can be created.

If it turns out that there exist non--distillable states in $\C^4\otimes\C^4$
with NPT as conjectured in \cite{Du99b,Di99a}, this automatically implies that
the conditions (i) for distillability obtained for the states $\rho_4$ are {\it
not} sufficient for arbitrary states $\rho$. To see this, let us consider the
question of distillability of a maximally entangled pair between $C$ and $D$.
Assume that the partial transposition with respect to the bipartite splits (2c),
(2d), (2f) and (2g) is negative. Let us concentrate on the bipartite split (2f).
According to the conjecture in \cite{Du99b,Di99a}, the negativity of the partial
transposition with respect to this split is not sufficient to ensure
distillability. So we have that there exist states which are not distillable
even if we allow for joint actions at $(AC)$ and $(BD)$ and are thus also not
distillable when allowing only local operations in $A,B,C$ and $D$ (As
mentioned earlier, each bipartite split provides us with an upper limit of what
can be done by local operations).

\section{Mixtures of GHZ-state with identity}\label{mixture}

We will apply now our results to the case in which we have a maximally entangled
state of $N$ particles mixed with the completely depolarized state
\be
\rho(x) = x |\Psi_0^+\rangle\langle \Psi_0^+| + \frac{1-x}{2^N} \eins.
\ee
This is clearly a special case of the state $\rho_N$ with
$\lambda_0^-=\lambda_j=\frac{1-x}{2^N}$, $\lambda_0^+=x+\frac{1-x}{2^N}$ and thus
$\Delta=x$. These states have been analyzed in the context of robustness of
entanglement \cite{Vi99}, NMR computation \cite{Br99}, and
multiparticle purification \cite{Mu98}. In all
these contexts bounds are given regarding the values of $x$ for which
$\rho(x)$ is separable or purificable. For example, in Refs.\
\cite{Br99,Vi99} they show that in the case $N=3$ if $x\le
1/(3+6\sqrt{2}),1/25$ then the state is separable, respectively.
In Ref. \cite{Mu98} it is shown that for $N=3$ if $x>0.32263$ then $\rho(x)$ is
distillable. Using our results we can state that $\rho(x)$ is fully
non--separable and distillable to a maximally entangled state iff
$x>1/(1+2^{N-1})$, and fully separable otherwise. Specializing this for
the case $N=3$ we obtain that for $x>1/5$ it is non--separable and
distillable \cite{Sc99}.

Note also that the purification procedure proposed in this work is
pretty unefficient compared to the two--step procedure proposed in
\cite{Mu98}, although it allows us to determine stronger bounds
for the value of $x$. However, one can slightly modify the
procedure proposed in \cite{Mu98} such that the protocols P1 and
P2 are no longer performed alternately as in the original version,
but rather in a specific (state dependent) order, e.g.
P1-P1-P1-P2-P1 etc. When doing so, we found by numerical
investigation (for $N=3$) that all states $\rho_3$ which are
purificable to a GHZ state using our procedure are also
purificable using the modified procedure of \cite{Mu98}, which
thus provides an efficient purification protocol for states of the
form $\rho_3$.

%--------------------------------------------------------------------------------

\section{Summary}\label{Summary}

In summary, we have proposed a classification of arbitrary multi--qubit systems.
For a family of states, we gave a full characterization of the
separability and distillability properties. These states play the role of Werner
states in these systems since any state can be reduced to such a form by
depolarization. Thus, our results provide sufficient conditions for
non--separability and distillability for general states.

We thank M. Lewenstein, S. Popescu, G. Vidal, P. Zoller and especially R. Tarrach for
discussions. This work was supported by the \"Osterreichischer Fonds zur F\"orderung
der wissenschaftlichen Forschung, the European Community under the TMR
network ERB--FMRX--CT96--0087, and the Institute for Quantum Information
GmbH.

% -------------------------------------------------------------

\narrowtext
\begin{table}
\begin{tabular}[t]{||c|l|l||}
 Positive Operators         & Class & Distillability                     \\ \hline
 None                       &  1    & (GHZ) $|\Psi_0^+\rangle_{ABC}$     \\ \hline
 $\rho_3^{T_A}$             &  2.1    & (Pair) $|\Phi^+\rangle_{BC}$       \\ \hline
 $\rho_3^{T_A},\rho_3^{T_B}$&  3.1    & Activate with $|\Phi^+\rangle_{AB}$\\ \hline
 All                        &  5    &                                    \\
\end{tabular}
\caption[]{Separability and distillability classification of $\rho_3$}
\label{Table1}
\end{table}

%%%%%%%%%%  here the figures  %%%%%%%%%%%%%%%%%%%%%%%%%%%%%%%%%%%%%

\end{document}